\documentclass[twocolumn,showpacs,amsmath,amssymb]{revtex4}

\usepackage{latexsym}
\usepackage{graphicx}
\usepackage{pifont}
\usepackage{color}
\usepackage{epsfig}
\newcommand{\beqn}{\begin{equation}}
\newcommand{\eeqn}{\end{equation}}
\newcommand{\req}[1]{Eq.\,(\ref{#1})}

\begin{document}
\title{Compact Ultradense Matter Impactors}
\author{Johann Rafelski$^{1}$}
\author{Lance Labun$^1$}
\author{Jeremiah Birrell$^2$}
\affiliation{$^1$Department of Physics, $^2$Program in Applied Mathematics, \\ The University of Arizona, Tucson, Arizona, 85721 USA}
\date{September 3, 2012 }

\begin{abstract}
We study interactions of meteorlike compact ultradense objects (CUDO), having nuclear or greater density, with Earth and other rocky bodies in the Solar System as a possible source of information about novel forms of matter.  We study the energy loss in {\small CUDO} puncture of the body and discuss  differences between regular matter and {\small CUDO} impacts.
\end{abstract}


\maketitle
{\bf  Introduction.}---%
What if there are ``dark'' matter meteor and asteroidlike bodies in the Universe?  Could some of them have collided with Solar System bodies and the Earth? On account of high density resulting in small geometric cross section and the high surface gravity, such {\bf c}ompact {\bf u}ltra{\bf d}ense {\bf o}bjects  ({\small CUDO}s)   ``dressed'' or not by normal matter are  likely   to survive  transit through Earth's atmosphere. {\small CUDO}s' high density of gravitating matter  provides  the distinct observable difference in the outcome of their collisions with rocky bodies, the surface-penetrating puncture: the {\small CUDO} will practically always enter the target body, and many will exit the body,   with only a fraction of the kinetic energy damaging the solid surface. We therefore refer to {\small CUDO} collisions with Solar System rocky bodies as ``punctures''.

We describe here  the physics of {\small CUDO} puncture focusing on Earth, and extend the considerations as applicable to Mars, the Moon and other solar rocky bodies. Simple features are discussed which distinguish their appearance from normal matter meteorite impacts. This should provide  a path to recognizing the presence of asteroid   {\small CUDO}s in the Universe, and as such offer new insights about novel  forms of  ultrahigh density matter, such as  strangeness-rich nuclear matter~\cite{Bodmer:1971we,Witten:1984rs}. Light nuclearite  strangelets have not been found  soil from  the Moon~\cite{Han:2009sj}, nor in laboratory production experiments~\cite{Abelev:2007zz} but the space search continues with ongoing efforts to observe the strangelet flux with the AMS-spectrometer mounted on the International Space Station~\cite{Sandweiss:2004bu}.

This work   extends the mass range of prior puncture studies. a) By  strange quark matter   ``nuclearites''~\cite{DeRujula:1984ig} of $M\lesssim 10^4m_{\rm proton}$ were recognized as passing across a rocky body on account of their tiny size and subatomic self-binding, and capable of generating  seismic disturbances~\cite{DeRujula:1984ig,Herrin:1995es}, such seismic signatures  have been searched for on Earth~\cite{Herrin:2005kb} and Moon~\cite{Banerdt:2006}; b) Similar consideration was also made for  microsized black holes~\cite{Jackson:1973},  that work was considerably extended recently~\cite{Khriplovich:2007ci,Luo:2012pp}.

The {\small CUDO} mass range that is not excluded by experiment is rather wide.  An upper mass limit for massive ``invisible'' objects is provided by microlensing surveys: massive compact halo objects ({\small MACHO}s) with $M\gtrsim M_{\oplus}=5.97\times10^{24}\:{\rm kg}$ (i.e. larger than  Earth's mass) are ruled out~\cite{Carr:2009jm,Massey:2010hh}. {\small MACHO}s comprise two object classes: (i) normal density objects  such as planets and dwarf stars, and (ii) neutron matter  and higher density objects which we study here. The lower mass limit originates in gravitational stability such as would arise from a relatively loose conglomerate of otherwise noninteracting heavy dark matter  particles.

We begin considering the potential sources of {\small CUDO} and the flux and puncture frequency by massive {\small CUDO}s. We then turn to a discussion of energy deposition on puncture. We close with a short discussion of the   event signatures. 

{\bf Potential sources of {\small CUDO}s  and {\small CUDO} rate of collisions.}---%
{\small CUDO} nuclearite  meteors can be produced as fragments from a collision of neutron star cores that contain strange quark matter (nearly $s=u=d$-quark abundance symmetric matter)~\cite{Madsen:2004vw,Bauswein:2008gx}. Such macroscopic nuggets of strange quark matter may be sufficiently bound to be stable on a cosmological time scale~\cite{Labun:2011yr}.  However, the expected flux of nuclearite {\small CUDO}s created in stellar evolution events is naturally small and their rocky body impacts would be very rare.

Even if the flux of massive {\small CUDO}s such as nuclearite strangelets decreases with their mass like $1/M$~\cite{DeRujula:1984ig,Madsen:1989pg}, consideration of the integrated flux over a period of billions of years  will result in the accumulated number of heavy puncture events  being  much more numerous than the expected short observation period acoustic signal events of  light {\small CUDO}s.  

During this long puncture ``observation'' time the Solar System samples a large  domain of the Milky Way, circling the galactic center a few times and, in particular, it passes through spiral arm regions comprising higher density of  visible matter. It is likely that  {\small CUDO} abundance of any type is enhanced in high visible matter density domains of the galaxy, resulting in periodic enhancement of {\small CUDO} flux within the solar system as compared to the present condition.

Study of the Bullet Cluster~\cite{Bullet2006} demonstrates dark matter as an independent, dynamical and gravitating component. There is no direct or indirect observational evidence constraining  the  state in which dark matter is to be found. The estimated dark matter density  in the local area of the Milky Way is $\rho_d=5.3\times 10^{-22}\:{\rm kg/m^3}$, roughly equal to the averaged local density of visible matter~\cite{PDG2010}. Overall, by mass, the dark matter is about 4 times as abundant as is visible matter.

Considering that dark matter is the dominant form of matter in our time, recent (on the scale of 100 millions of years) impacts of dark matter {\small CUDO}s should be  frequent: The Sun and the large Solar System bodies comprise  a vastly dominant fraction  of the visible matter comoving with Earth in the galaxy. Since by mass the abundance of visible and dark matter is similar, it follows that   even if a relatively small fraction e.g. $10^{-6}$  of dark matter is bound in meteor-sized objects, their flux would  be similar to the number of normal matter meteorites, though the distribution in space could be very different.

The above argument presumes that  a tiny fraction of all dark matter is clumped. Clumping of the very heavy ``cold'' dark matter dust has so far received little attention and a  study of this process is beyond the scope of this work. However, large scale visible matter structure in the Universe depends on the existence of gravitating dark matter density fluctuations formed in the early Universe.  Therefore the question is not if dark matter fluctuations exist but if yet higher density gravitationally bound dark matter objects can arise.

Beyond the first step described above there could be a continued  mutual speed-up and reinforcement of visible matter and  dark evolutionary dynamics: once enough visible matter converged together near  the first dark matter fluctuations, and first stars form, these stars become  themselves gravitational attractors helping to aggregate  dark matter, assisting in the formation of dark gravitationally self bound objects, which can be freed  at the end of  some stellar evolution cycles.  The lighter {\small CUDO}s with a mass significantly below that of the Earth  are those that would be of interest in the puncture process we address here.

Like in the study of neutron stars, dark matter {\small CUDO} properties are obtained solving Oppenheimer-Volkov   equations   for a heavy  gravitating dust~\cite{Narain:2006kx,Dietl:2011cs}. The higher the mass of the cold dark particles, the smaller   the  Oppenheimer-Volkov  upper mass stability limit, and the fewer particles that are needed to form self-bound body.  Should the upper  stability mass limit be exceeded, a dark black hole is formed in a gravitational collapse~\cite{Luo:2012pp}. Aside from this upper mass limit there is a  minimum mass  required for the presence of sufficiently high self gravity to provide  stability in during the Universe's evolution to the current epoch.

We  consider punctures  by  a member of the cloud  of  objects comoving with the solar system.  Any such accompanying {\small CUDO}s have isotropic velocities $v_{\rm gal}\sim$ a few km/s relative to the solar system.  Given that the two classes of collisions (normal and {\small CUDO}) with rocky bodies were not distinguished before, we consider the velocity distribution of  {\small CUDO}s  to be similar to that of nonsolar system visible matter impactors.  We note a potential third class, a fast {\small CUDO}   impactor: should {\small CUDO}s constitute a fraction of the primordial dark matter, they should form an independently rotating galactic halo, resulting in a much higher relative velocities   $v_{\rm halo}\sim 200\:{\rm km/s}$. Seeing  the inverse dependence on $v$ in the equations below, such impacts will be yet less damaging to the target body, but entail a much greater local force.

A conservative {\small CUDO} puncture rate is obtained by ignoring the concentration of {\small CUDO}s in the proximity of the Solar system, i.e. spreading {\small CUDO} matter, taken as equal to the solar system visible matter, over the volume cell in the Milky Way occupied by the solar system.  For comoving {\small CUDO}s the gravity of the Sun focuses objects passing nearby, increasing the flux of objects whose trajectories enter the inner solar system (achieving a minimum radius smaller than  Earth's orbit $R_{\rm orbit,E}$) and enhancing in that way the effective  cross section  
\beqn\label{sigmaeff}
\sigma 
= \left(1+\frac{2GM_{\odot}}{v_{\rm gal}^2R_{\rm orbit,E}}\right)4\pi R_{\oplus}^2.
\eeqn 
($4\pi R_{\oplus}^2$ is the planetary detector surface area.)

We do not know the {\small CUDO} mass distribution in the flux. Hence what follows is for an average mass compact impactor. The {\small CUDO} number density  $n=\rho_d/M$, leads to\,an\,expected\,annual\,event\,rate\,for\,puncture\,on\,the\,Earth, 
\beqn\label{galacticflux}
N_{\oplus}(v,M) = n\sigma v 
 \simeq 3\: 10^{-18} \frac{\rm km/s}{v}\frac{M_{\oplus}}{M}\:{\rm yr^{-1}}
\eeqn
where $M$ is the mass of a typical {\small CUDO} and $v$ is the velocity attained by the object at $R_{\rm orbit,E}$.  We expect up to a dozen collisions with {\small CUDO}s of $M<10^{-9}M_{\oplus}$ over the course of Earth's history.  If the {\small CUDO} mass spectrum parallels the normal matter impactor mass spectrum~\cite{Stuart04}, lower mass punctures will be significantly more frequent and collisions with $M<10^{-15}M_{\oplus}$ (i.e. objects of mass below 6 million tons) occurring as often as every few hundred years.

{\bf {\small CUDO} punctures.}---%
A {\small CUDO} comoving at low relative velocity outside the solar system achieves $v\sim 20-60\:{\rm km/s}$ relative to Earth.  As a result, the {\small CUDO} penetrating the crust will be supersonic in the mantle ($c_s \sim 8\:{\rm km/s}$).  Understanding the dynamics of a supersonic compact gravitating body passing through a dense planetary medium is the key physics challenge.  We generalize and expand on preceding considerations of punctures by micro black holes~\cite{Jackson:1973,Khriplovich:2007ci}.

We propose estimates of magnitudes based on linear response and two mechanisms of energy exchange with planetary matter:
(i) pulverization and entrainment of rock nearest the trajectory of the object, and
(ii) creation of a supersonic shock.  
We expect the assumption of linear response  breaks down for masses greater than some $M_s(v)$, resulting in a more inelastic nonlinear response.  The following scaling considerations can be considered valid only for $M<M_s$ and further exploration of $M_s(v) <10^{-2}M_{\oplus}$ is left to future considerations.

\begin{figure}
\includegraphics[width=0.45\textwidth]{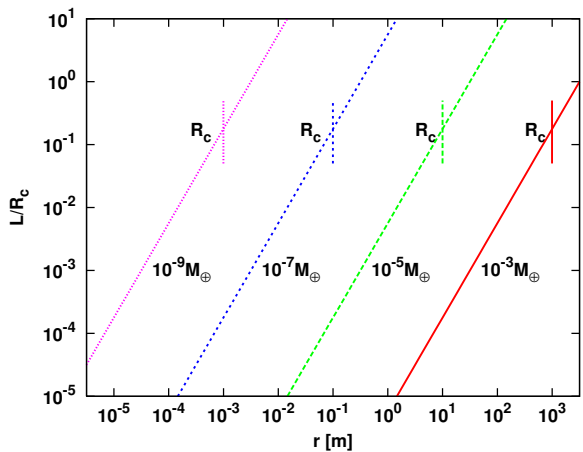}
\caption{ The tidal fracture length scale $L$ from Eq.\,(\ref{fracture}) normalized to the capture radius $R_c$ Eq.\,(\ref{Rcap}) as a function distance from the trajectory.  The velocity of the compact object is fixed at $v=40\:{\rm km/s}$ and $c_s=8\:{\rm km/s}$ used for the speed of sound in the mantle; a higher $v$ only reduces $L/R_c$.  
\label{fig:fracleng}}
\end{figure}

{\it Entrainment of material.}---%
In order to carry material away, the {\small CUDO} tidal force must break it into pieces small enough to be gravitationally captured.  When the differential (tidal) acceleration over a length $L$ exceeds the capability of the material to change density elastically, the integrity of the material is compromised, we expect a fracture to open  within the length $L$.  We compare the differential gravitational pressure, which is the product of the tidal acceleration and mass per unit area,
\beqn
P(r-L/2)-P(r+L/2)=\frac{2GML}{r^3}\rho L,
\eeqn
to the material's bulk modulus $K=\rho c_s^2$, which is a function of the density $\rho$ and the sound speed $c_s$.  $L$ then scales with the distance $r$ from the trajectory as
\beqn\label{fracture}
\frac{L}{R_c} = A_f \left(\frac{r}{R_c}\right)^{3/2}.
\eeqn
with $A_f = \sqrt{2}(c_s/v)$. $R_c$ is the capture radius~\cite{Zeldovich:1971}
\beqn\label{Rcap}
R_c:=\frac{4GM}{v^2} 
= 10^6\left(\frac{40\:{\rm  km/s}}{v}\right)^2\frac{M}{M_{\oplus}}\:{\rm m}
\eeqn
within which material can be entrained with the object and acquires its velocity $v$.  Figure~\ref{fig:fracleng} shows that $L/R_c<0.1$ for radii less than $R_c$.  All capture-eligible material is broken into pieces small enough to be pulled along with the object, thus forming a slug of fine particulate matter carried with the {\small CUDO} even as it exits the target body.

The velocity of the {\small CUDO} is reduced by the capture of  material, because the slug of entrained material must acquire the comoving kinetic energy, the average gain in potential gravitational energy being small in comparison
\beqn \label{KEcaptured} 
\Delta E 
 = A_k \rho R_{\oplus}\pi R_c^2\frac{v^2}{2}
 \simeq 6.4\times 10^{31} \left(\frac{40\:{\rm km/s}}{v}\right)^{\!\!2}
                        \left(\frac{M}{M_{\oplus}}\right)^{\!\!2}\:{\rm J}
\eeqn 
$A_k$ contains factors for capture efficiency (which may differ from unity due to drag between captured and neighboring material) and impact parameter $b$ (which determines total volume of tube radius $R_c$ around trajectory).  In an ideal fluid model, capture efficiency is 100\% and $A_k=2\sqrt{1-b^2/R_{\oplus}^2}$.  The numerical coefficient is obtained with the density of mantle $\rho\simeq 4000\:{\rm kg/m^2}$ and considering $A_k=1$. However, there is also  the kinetic energy and the drag of uncaptured material which we expect to be  of similar magnitude.

Comparison to the kinetic energy  before collision,
\beqn\label{fracloss}
\frac{\Delta E}{E}
 = A_k\frac{\rho }{M}R_{\oplus}\pi R_c^2
 \simeq 0.013\left(\frac{40\:{\rm km/s}}{v}\right)^4\frac{M}{M_{\oplus}}
\eeqn
suggests that objects $M<10^{-4}M_{\oplus}$ pass completely through  Earth without being stopped and the energy transferred to the entrained matter is at and below 1/1000. While the total energy deposited in the target body is a small fraction of the total kinetic energy, the energy deposited by the {\small CUDO} at the surface is still much smaller.  A normal matter meteorite is stopped soon after impacting the surface, which requires that it dissipates all of its kinetic energy in a relatively small target area.   Therefore the {\small CUDO} puncture is much less destructive than  normal matter meteorite.

We note features specific to the case of strangelet {\small CUDO}s which originate in the electrical charge of the strange quark matter core. Strangelet {\small CUDO}s  are expected to  acquire a dressing of normal matter, forming a ``crust" \cite{Alcock:1986hz} .  For the  mass range of interest, the kinetic energy will suffice to press additional matter across the electric potential into the quark matter core.  Consequently, additional matter can be captured. The resulting change in strangelet mass will depend primarily on the amount of matter encountered during the puncture: we find the fractional mass change  corresponding to \req{fracloss} is of order $10^{-3}(M/M_{\oplus})$,  and only 150 times larger when considering the Sun as a target. However, a charged strangelet {\small CUDO}~\cite{Abers:2007ji} would lose additional energy during the puncture due to electrical ionization and polarization effects, which may be particularly important in strangelet collisions with the Sun.   Solar impacts as well as strangelet-strangelet collisions  depend further on strange quark matter bulk properties. These are the subject of present day interest  in view of nontrivial color-superconducting and color-flavor locked cold quark matter phases~\cite{Alford:2007xm}. 

To establish that material in and around the slug is indeed molten, we look at temperature in the bow shock
\beqn
k_B T=A_T v^2
\eeqn
with $A_T$ depending on the atomic composition of the material.  Taking an ideal gas model~\cite{Ruderman:1971} in view of the high temperatures expected gives $A_T=3\times 10^{-27}\:{\rm kg}$ and $T\simeq 10^5\:{\rm K}$.  Estimating based on the continental secular cooling rate of Earth $\sim 80\:{\rm mW/m^2}$, the conduit formed by the puncture could remain a significant local source of thermal energy, recognized by transient  deep-origin volcanism not related to plate boundaries. 

{\it Supersonic shock wave.}---%
At ranges larger than $R_c$ tidal stresses generate an intense seismic wave.  Since the velocity of the {\small CUDO} is greater than the speed of sound in the mantle or crust, the wave is compressed into a shock. Energy deposition via coherent acoustic emission has been given as  Eq.\,(13) in \cite{Khriplovich:2007ci} 
\beqn
\left.\frac{dE}{dx}\right|_{\rm shock}\!\!\!\!\!
=A_s \pi R_c^2\rho v^2 \simeq 1.9\times 10^{26}\, \left(\frac{M}{M_{\oplus}}\right)^{\!\!2}\left(\frac{\rm 40\:km/s}{v}\right)^{\!\!2}\:\frac{\rm J}{\rm m}
\eeqn
where $A_s=(1/8)\ln(c_s^2/4\pi G\rho a^2)$, $a$ being typical inter atomic spacing. This estimate finds a large amount energy in the compression wave, 40 times the captured matter kinetic energy Eq.\,(\ref{KEcaptured}). 

{\bf {\small CUDO} stability on impact.}---%
Micro black holes are stable and puncture through the target. Nuclearite {\small CUDO}s are bound by nuclear scale forces and if stable on cosmological time scale,  atomic scale forces cannot pull them apart. On account of the much larger dark matter particle mass, the dark matter  {\small CUDO} has a very high  density, and its Roche limit ($2\rho_t<\rho_c$ for {\small CUDO} at target surface) is not relevant in  stability consideration. In fact our evaluation of the effect of a {\small CUDO} on the target body is like an ``inverse'' Roche process in that the target  structure is locally broken by the tidal forces of the high density {\small CUDO}.
 
Conversely, for a gravitationally self-bound {\small CUDO} of sufficiently small mass, the potential energy at the surface will be small enough to allow the target induced  polarization force to shear and attract particles from the {\small CUDO}.  The qualitative condition for transfer of matter is that the presence of the target body opens a potential valley from the binding potential of the {\small CUDO} at its surface $R_c$ towards the potential of the target  body `$t$'. The minimal stable mass condition arises from the requirement that transfer of {\small CUDO} particles to target begins  after surface penetration,
\beqn\label{transfer}
M_c>M_t\frac{R_c}{R_t}.
\eeqn
We find a large domain between this lower, and the  upper gravitational collapse limit,  wherein one can consider  {\small CUDO} collisions with Earth and other rocky bodies~\cite{RDL:2012}.

{\bf {\small CUDO}s in Particle Physics.}---%
The features of {\small CUDO} impacts discussed here arise from their
compactness, and the mass-radius scale, for beyond-standard-model ({\small
BSM}) {\small CUDO}s  set by the energy-mass scale of the dark matter
particles~\cite{Narain:2006kx,Dietl:2011cs}.  Strangelet {\small CUDO}s
and  {\small BSM} {\small CUDO}s are distinguished by the large splitting 
between their respective energy scales, $\sim 1~{\rm GeV}$ and $\gtrsim
250~{\rm GeV}$. Terrestrial impacts should allow differentiation on the
corresponding scale of impacting {\small CUDO}s mass and size.
Considering the upper and lower mass limits originating in the
constituent particle properties~\cite{RDL:2012}, a constraint on the
dark matter energy-mass scale could arise. We expect to at least  be
able to differentiate heavy BSM particle {\small CUDO}  from e.g.
color-flavor-locked quark matter~\cite{Alford:2007xm,Franzon:2012in}.
Such quark matter {\small CUDO} impacts are believed to have several
distinctive features~\cite{Paulucci:2009zz}.

Another aspect controlled by the  {\small BSM}-dark matter energy scale
is {\small CUDO} formation scenario. Compact, and relatively small in 
particle number {\small CUDO}s could be formed primordially by density
fluctuations~\cite{Hansson:2004km}, potentially absorbing a large fraction 
of dark matter particles as used in flux estimate \req{galacticflux}.
Strangelet {\small CUDO}s can be formed in more recent times, e.g. in
collapsing stars~\cite{Bauswein:2008gx}.

{\bf  Discussion and outlook.}---%
The salient observable feature of {\small CUDO}s is their capability to create puncture in rocky bodies. On account of the relatively small energy loss distributed across the puncture conduit, even relatively massive {\small CUDO} punctures preserve  the integrity of the target and inflict in such event a relatively small shotlike damage. 

While the exit ``wounds'' must be fewer than puncture entries, on account of possible {\small CUDO} instability   after entry,  their appearance should be  more characteristic considering that there is little else in the Universe that can generate an exit conduit from a body and, considering  Earth puncture, can transport material into the upper atmosphere.  A puncture event  may be thus recognized by coincidental presence of impact and ultrastrong ``volcanic'' eruption. The AD 536 event~\cite{Rigby:2004,Larsen08,Ferris11} which has lead to a major historical climate excursion  invites reexamination as a possible full {\small CUDO} puncture, i.e. impact and  exit accompanied by material transport into the upper atmosphere.

Most recent and present epoch {\small CUDO}s are possibly dressed in normal matter and thus many punctures are also ``normal'' meteorite impacts.  Depending on the ratio of normal to {\small CUDO} mass and the composition of normal matter component, there are characteristic  {\small CUDO} features in comparison to impacts by solely normal meteors:  (a) Preceding the impact, in transit through atmosphere, the {\small CUDO} core binds and stabilizes even the heated  meteorite material, making the normal matter impactor appear exceptionally stable. (b) At impact on the surface, the normal matter material will largely evaporate, if not entrained with the {\small CUDO} into the mantle and beyond.  A possible  signature of {\small CUDO} surface impacts is then the absence of both  {\small CUDO} and normal impactor material. (c)  At exit, the kinetic energy of the moving {\small CUDO} may create an exit ``hump'' and/or appearance of lava flow in environments where none should be present.  While rapid terrestrial surface evolution would mask some exit features, they should be well preserved on Mars, the Moon, and other rocky bodies. 

Present day  models of meteor impacts do not introduce the possibility of a high density impactor and the resulting surface puncture. It is the absence of the CUDO type option in detailed impactor models that in our opinion leads to ongoing intense discussions  within diverse geological and planetary communities regarding the priority of impacts versus eruptions, raising further questions about the  mechanism of major  unexplained phenomena (e.g., ``Volcanism, impact and mass extinctions: incredible or credible coincidences?''~\cite{White:2005}). These topics are well beyond the current context, yet the wide ranging debates are indicating the  need for a novel event type, and   puncture seems to fit the need.

Research  supported in part by  the  U.S. Department of Energy, Grant No. DE-FG02-04ER41318 and by the Department of Defense (DOD) through the National Defense Science \& Engineering Graduate Fellowship (NDSEG) Program.


\end{document}